\documentclass[10pt,preprint2]{aastex}

\def\figdir{.}

\def\HII{\mbox{\ion{H}{2}}}

\def\dim#1{\mbox{\,#1}}
\def\figname#1{\figdir/#1}

\begin{document}

\pagestyle{myheadings}
\markright{DRAFT: \today\hfill}

\title{The Damping Wing of the Gunn-Peterson Absorption and Lyman-Alpha
  Emitters in the Pre-Reionization Era}

\author{Nickolay Y.\ Gnedin\altaffilmark{1,3} and Francisco
  Prada\altaffilmark{2,3}} 
\altaffiltext{1}{CASA, University of Colorado, Boulder, CO 80309, USA;
gnedin@casa.colorado.edu}
\altaffiltext{2}{Ram\'on y Cajal Fellow, Instituto de Astrofisica de
  Andalucia (CSIC), E-18008 Granada, Spain; fprada@iaa.es}
\altaffiltext{3}{Visitor at the Kavli Institute for Cosmological Physics,
  University of Chicago}

\begin{abstract}
We use a numerical simulation of cosmological reionization to estimate the
likelihood of detecting Lyman-$\alpha$ emitting galaxies during the
pre-reionization era. We show that it is possible to find galaxies even at
$z\sim9$ that are barely affected by the dumping wing of the Gunn-Peterson
absorption from the neutral IGM outside of their $\HII$ regions. The
damping wing becomes rapidly more significant at $z>9$, but even at $z>10$
is it not inconceivable (although quite hard) to see a Lyman-$\alpha$
emission line from a star-forming galaxy.
\end{abstract}


\section{Introduction}

A recent landslide in the number of detections of high redshift
Lyman-$\alpha$ emitters (Dey et al.\ 1998; Hu, Cowie, \& McMahon 1998;
Weymann et al.\ 1998; Hu, McMahon, \& Cowie 1999; Stern \& Spinrad, 1999;
Ellis et al.\ 2001; Ajiki et al.\ 2002; Hu et al.\ 2002; Kodaira et al.\
2003; Taniguchi et al.\ 2003; Rhoads et al.\ 2003; Maier et al.\ 2003; Cuby
et al.\ 2003; Bunker et al. 2003; Hu et al.\ 2004; Pello et al.\ 2004;
Kneib et al.\ 2004) offers a unique probe of the ionization history of the
universe during cosmological reionization. As observations of
Lyman-$\alpha$ absorption in the spectra of high redshift quasars (Djorgovski
et al.\ 2001; Becker et al.\ 2001; Fan et a.\ 2002, 2003; White et al.\
2003; Songaila 2004) unambiguously indicate that the universe was reionized
shortly before $z=6$ (Gnedin 2002), the observations of Lyman-$\alpha$ at
$z>6$ directly probe the pre-reionization era.

Specifically, since the damping wing of the Gunn-Peterson absorption (Gunn
\& Peterson 1965) of the neutral IGM will affect the Lyman-$\alpha$
emission line if the $\HII$ region around the emitting galaxy is not large
enough (Miralda-Escud\'e 1998; Haiman 2002), the observations of high
redshift Lyman-$\alpha$ emitters tell us something about the distribution
and sizes of cosmological $\HII$ regions - although, to understand what
specific constraints the observations place is a horrendously difficult
task\footnote{Since most Lyman-$\alpha$ emitters are discovered because
  they are highly lensed by foreground clusters of galaxies, determining
  the observational selection function is extremely difficult.}.

However, the very fact of discovering a high-redshift Lyman-$\alpha$
emitter may 
potentially be used to constraint the ionization history of the
universe. For example, Pello et al.\ (2004) report a tentative discovery of
a $z=10$ galaxy. Is it possible to see a Lyman-$\alpha$ emitting galaxy at
such a high redshift? Loeb et al.\ (2004) used this fact to put some
preliminary constraints on the ionization state of the IGM around that
galaxy. While a real situation is likely to be more complicated because of
the variety of reasons ($\HII$ regions are typically highly asymmetric
because I-fronts propagate much faster across underdense voids then across
dense filaments; bright galaxies are biased, so it is likely that more than
one galaxy is located inside a single $\HII$ region; Lyman-$\alpha$
emitters can also be located inside $\HII$ regions of luminous quasars,
which are often many times larger than $\HII$ regions of galaxies, etc), it
is interesting to investigate how likely it is to find a high redshift
galaxy unaffected by the damping wing of the Gunn-Peterson absorption from
the surrounding IGM.

\section{Star Formation Modeling}

In order to construct a model for the Gunn-Peterson absorption in the
pre-reionization era, we use a numerical simulation of cosmological
reionization similar to the one described in Gnedin (2000). The new
simulation, however, incorporates a significantly more accurate method for
following the time-dependent and spatially-inhomogeneous radiative transfer
using the Optically Thin Variable Eddington Tensor (OTVET) approximation of
Gnedin \& Abel (2001). 

The simulation assumes a flat cosmology with the values of cosmological
parameters as measured by the {\it WMAP\/} satellite (Spergel et al.\
2003)\footnote{Specifically, we assume $\Omega_m=0.27$, $\Omega_b=0.04$,
  $h=0.71$, $n=1$, and $\sigma_8=0.85$.}. 
The size of the simulation volume is $8h^{-1}\dim{Mpc}$ in comoving
units, and the comoving spatial resolution of the simulation is
$2h^{-1}\dim{kpc}$. The simulation included $128^3$ dark matter particles
(with mass of $2.6\times10^7M_\odot$), an equal number of baryonic cells on a
quasi-Lagrangian moving mesh, and about 3 million stellar particles that
formed continuously during the simulation. Only PopII stars are included in
the simulation as sources of ionizing radiation.

The simulation was adjusted to reproduce the observed evolution of the mean
transmitted flux in the Lyman-$\alpha$ line between $z=5$ and $z=6.4$
(White et al.\ 2003; Songaila 2004). We, therefore, can be confident that
the evolution of the IGM is treated correctly in the simulation at least at
$z\sim6$. 

Given a simulation, we identify all star-forming galaxies using the
DENMAX halo finding algorithm of Bertschinger \& Gelb (1991). For each
galaxy, we generate 100 lines of sight that originate at the galaxy
location and go along randomly chosen directions. We then generate an
absorption spectrum along each line of sight following the standard
procedure. For the computational efficiency, we use Liu et al.\ (2001)
approximation for the Voigt profile.

\begin{figure}[t]
\plotone{\figname{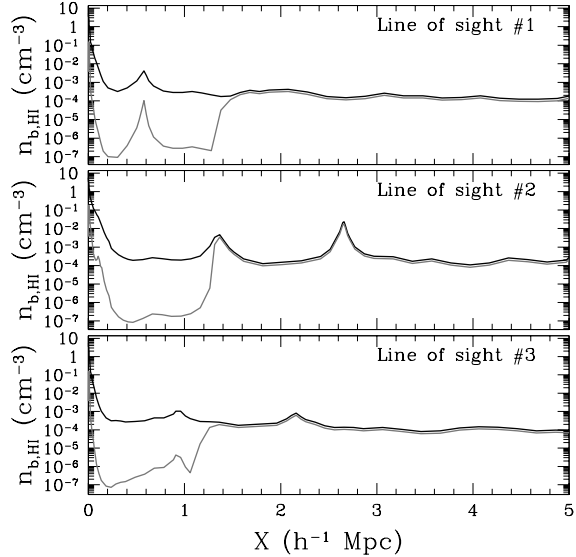}}
\caption{Baryon number density (black lines) and neutral hydrogen number
  density (gray lines) along 3 random lines of sight originating at the
  location of the most luminous galaxy at $z=9$ in our simulation.}
\label{figLR}
\end{figure}
Figure \ref{figLR} illustrates the distribution of the baryon number
density and neutral hydrogen number density along 3 random lines of sight
originating at the location of the most luminous galaxy at $z=9$. The
$\HII$ region around the central galaxies is easily visible. Because the
brightest 
galaxies are highly clustered, smaller galaxies and other dense structures
are often located within the $\HII$ regions of brighter ones. For example,
in the top panel, an overdensity of the order of 10 is located within the
$\HII$ region of the central galaxy. This overdensity is the outer part of a
smaller galaxy, and is, therefore, significantly more neutral than the rest
of the $\HII$ region (in fact, more neutral than simple ionization
equilibrium would predict, due to shadowing and time-dependent
ionization). This neutral spike is sufficient to produce large damping wing
absorption at the location of the galaxy and make the Lyman-$\alpha$ of the
central galaxy unobservable. Two bottom panels, however, show different,
``clean'' lines of sight that are not obscured by neighbors and produce
lower absorption by the damping wing of the neutral IGM outside the $\HII$
region. This illustrates that the effect of the damping wing of the
Gunn-Peterson absorption strongly depends on the direction of the line of
sight: the same Lyman-$\alpha$ emitting galaxy may be observable or
non-observable depending on the direction of viewing.

A significant complication, however, for the evaluation of the strength of
the Lyman-$\alpha$ emission line from the central galaxy is the galaxy
itself. Synthetic Lyman-$\alpha$ spectra from simulation are often treated
as {\it absorption\/} spectra, when, in reality, they are {\it
  scattering\/} spectra. For gas along a line of sight well outside the
source of emission this is an appropriate approximation, since scattering
off the direction of viewing removes the photons that could be observed,
and, thus, appears as effective absorption. This is not true for the source
of emission, since photons that were originally emitted in directions
different from the line of sight are scattered into the direction of
viewing. Thus, without the proper line radiative transfer treatment (which
is not included in our simulation), it is not possible to accurately
predict shapes of the Lyman-$\alpha$ emission lines from simulated
galaxies. Leaving such treatment for future work, in this paper we are
concerned only with an-order-of-magnitude estimate of the effect
that damping wings of the neutral gas outside $\HII$ regions have on
Lyman-$\alpha$ emission lines. 

In order to obtain such an estimate, we adopt a following simple
procedure. We assume that the Lyman-$\alpha$ emission line from a typical
galaxy has a width of the order of $150-200\dim{km/s}$. Thus, we exclude
all Lyman-$\alpha$ absorption within a specified velocity distance $v_{\rm
  off}$ from the systemic velocity of a simulated galaxy (because within
this distance we should treat the Lyman-$\alpha$ as scattering and not
as absorption). Also, since the significant suppression of Lyman-$\alpha$
emission at the systemic velocity may still make the red wing of the
emission line visible, we choose the Lyman-$\alpha$ optical depth at
$-150\dim{km/s}$ as a quantity that roughly characterizes the damping wing
absorption of the Lyman-$\alpha$ emission line. Typical Lyman-$\alpha$
emission lines have gaussian widths of the order of $60-80\dim{km/s}$, so
$-150\dim{km/s}$ roughly corresponds to $2\sigma$ to the red of the line
systemic velocity. In this paper we use a range
of values for $v_{\rm off}$ to estimate the sensitivity of our results to
that somewhat arbitrary parameter.

With this approach, we can calculate the Lyman-$\alpha$ optical depth at
$-150\dim{km/s}$ for each of 100 lines of sight for each of the simulated
galaxy. Thus, there are 100 values of the damping wing absorption optical
depth $\tau_{\rm Ly\alpha}$, or, equivalently, the flux decrement $F_{\rm
  Ly\alpha}=\exp(-\tau_{\rm Ly\alpha})$, for each galaxy. These 100 values
are not all the same, of course, so we can use them to approximately
characterize the probability that a given galaxy (labeled by its star
formation rate) has a given value of the Lyman-$\alpha$ flux decrement.

\section{Results}

\begin{figure}[t]
\plotone{\figname{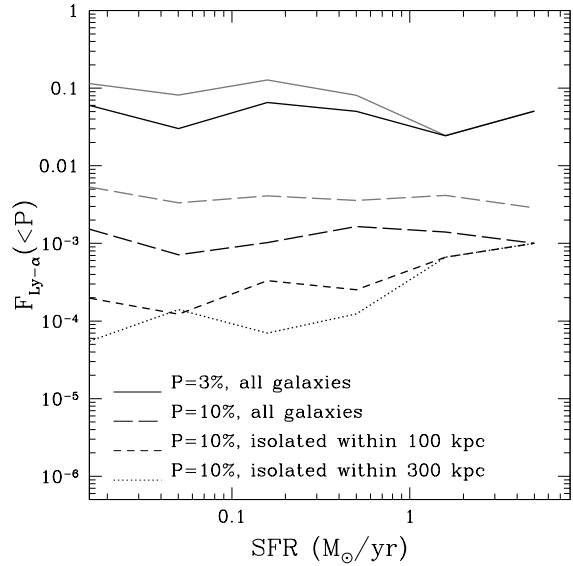}}
\caption{The limiting Lyman-$\alpha$ flux decrement as a function of star
  formation rate of a central galaxy for top 3\% (solid lines) and top 10\%
  (long-dashed lines) of all galaxies. Black lines show the case of $v_{\rm
  off}=150\dim{km/s}$, and gray lines show the case of $v_{\rm
  off}=200\dim{km/s}$. The short-dashed and dotted lines show the the
  limiting Lyman-$\alpha$ flux decrement as a function of star formation
  rate for isolated galaxies with $100\dim{kpc}$ and $300\dim{kpc}$ (in
  physical units) exclusion radius respectively (see text for more
  details).} 
\label{figFS}
\end{figure}
As a test of our approach, we show in Figure \ref{figFS} the values for the
Lyman-$\alpha$ flux decrement above which 3\% and 10\% of all galaxies
lie - for example, solid lines indicate that for 3\% of all galaxies the
decrease in flux at $-150\dim{km/s}$ is not more than a factor of 10. The
two values for the parameter $v_{\rm off}$ are used - the difference
between black and gray lines roughly indicates the uncertainty of our
estimate. 

An interesting feature of Fig.\ \ref{figFS} is the lack of dependence of the
flux decrement on the star formation rate. Naively, one would expect that
galaxies with lower star formation rates create smaller $\HII$ regions
around them, and, therefore, should exhibit higher suppression of their
Lyman-$\alpha$ emission lines - clearly at odds with our result. However,
this naive argument misses an important fact that high redshift galaxies
are highly clustered, so it is likely for a faint galaxy to be located
close to a bright galaxy, within the bright galaxy $\HII$ region. In order
to test that this effect is responsible for the lack of dependence of the
flux decrement on the star formation rate, we compute the flux decrement
versus the star formation rate for the sub-sample of isolated galaxies. We
define a galaxy to be isolated if it does not contain a galaxy with a
higher star formation rate within a sphere of predefined radius $r_{\rm
  isol}$. Short-dashed and dotted lines in in Fig.\ \ref{figFS} show the
flux decrement for the top 10\% of all galaxies (an analog of the
long-dashed line) for sub-samples of isolated galaxies with $r_{\rm
  isol}=100\dim{kpc}$ and $300\dim{kpc}$ (in physical units)
respectively. As one can see, isolated galaxies exhibit a strong dependence
on the star formation rate for rates above about $0.1M_\odot/{\rm yr}$. At
even lower star formation rates a different effect takes over: since the
star formation in dwarf galaxies is highly episodic (c.f.\ Ricotti, Gnedin,
\& Shull 2003), the instantaneous value of the star formation rate and the 
size of an $\HII$ region cease to correlate.

\begin{figure}[t]
\plotone{\figname{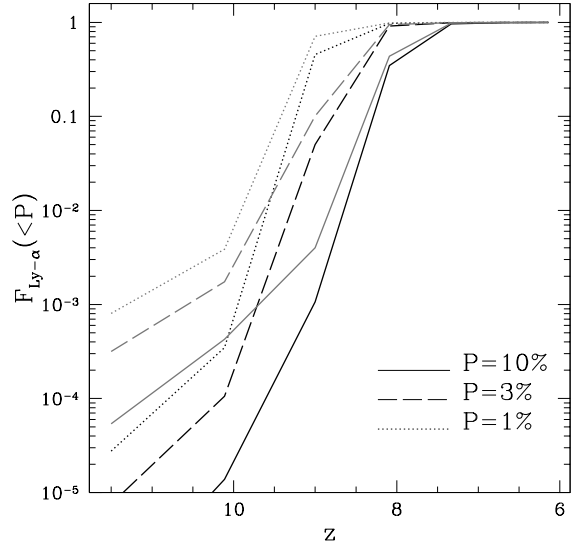}}
\caption{The limiting Lyman-$\alpha$ flux decrement as a function of star
  formation rate of a central galaxy for top 1\% (dotted lines), 3\%
  (dashed lines), and top 10\% (solid lines) of all galaxies as a function
  of redshift. Gray lines show the case of $v_{\rm
  off}=200\dim{km/s}$, and black lines show the case of $v_{\rm
  off}=150\dim{km/s}$ - the difference between the corresponding lines
  indicates the uncertainty of our calculations.}
\label{figFZ}
\end{figure}
The redshift dependence of the Lyman-$\alpha$ flux decrement is shown in
Figure \ref{figFZ}. As once can see, our estimates become highly uncertain
at $z\sim10$ - at these redshifts full radiative transfer in
Lyman-$\alpha$ line is required to make an accurate prediction for the
effect of the damping wing on the profile of the Lyman-$\alpha$ emission
line. At lower redshifts, however, our estimates are more precise.

\section{Discussion}

Somewhat unexpectedly, even at as high a redshift as $z\sim9$, one percent of
all Lyman-$\alpha$ emitters should have their emission lines barely
affected by the damping wing. At higher redshifts the damping wing
quickly becomes more significant, but finding Lyman-$\alpha$ emitters at
$z>10$ should still be possible, although they must be quite rare. 

The reason for this rapid transition is quite simple: as the universe
expands, $\HII$ regions increase in size, so at each moment there exists a
characteristic scale for the size distribution of $\HII$ regions (say, a
typical size of an $\HII$ region around an $L_*$ galaxy). At the same time,
the damping wing of the Gunn-Peterson absorption introduces another spatial
scale (about $\Delta z=0.01$ or proper distance of $200\dim{kpc}$ at
$z\sim10$; Miralda-Escid\'e 1998). Thus, when typical $\HII$ regions around
bright galaxies exceed this scale, a substantial fraction of all
Lyman-$\alpha$ emitters becomes observable.

It is important to emphasize here that our estimate is likely only a lower
limit: because of the finite size of the computational box, we miss the
brightest galaxies which create the largest $\HII$ regions (although the
size of the simulation box was selected so as to make this effect not
significant at $z>6$). In addition, our simulation does not include
quasars, which can create $\HII$ regions several times that of
galaxies. High redshift quasars are rare (there should be only about 2
SLOAN quasars in the whole observable universe between redshifts 9 and 10,
Fan et al.\ 2001), so that effect is most likely not large too, but
without a detailed simulation we cannot estimate it, so we treat our
results as a lower limit.

\acknowledgements
We thank A.\ Kravtsov for valuable comments and the Kavli Institute for 
Cosmological
Physics at the University of Chicago for hospitality during the time this
paper was initiated. 
This work was supported in part by NSF grant AST-0134373 and by
National Computational Science Alliance under grant AST-020018N and
utilized IBM P690 array at the National Center
for Supercomputing Applications.


\begin{references}

\reference{}
Ajiki, M., et al. 2002, ApJ, 576, L25

\reference{}
Becker, R.\ H., et al. 2001, AJ, 122, 2850

\reference{BG91}
Bertschinger, E., \& Gelb, J. 1991, J.\ Comput.\ Phys., 5, 164

\reference{}
Bunker, A.\ J., Stanway, E.\ R., Ellis, R.\ S., McMahon, R.\ G., \& McCarthy,
P.\ J. 2003, MNRAS, 342, L47

\reference{}
Cuby, J.\ G., LeFe`vre, O., McCracken, H., Cuillandre, J.-C., Magnier, E., \&
Meneux, B. 2003, A\&A, 405, L19

\reference{}
Dey, A., Spinrad, H., Stern, D., Graham, J.\ R., \& ChaAee, F.\ H. 1998, ApJ,
498, L93 

\reference{}
Djorgovski, S.\ G., Castro, S.\ M., Stern, D., \& Mahabal, A. 2001, ApJL,
560, 5 

\reference{}
Ellis, R., Santos, M.\ R., Kneib, J.-P., \& Kuijken, K. 2001, ApJ, 560, L119

\reference{}
Fan, X., et al. 2001, AJ, 122, 2833

\reference{}
Fan, X., Narayanan, V.\ K., Strauss, M.\ A., White, R.\ L., Becker, R.\ H.,
Pentericci, L., \& Rix, H.-W. 2002, AJ, 123, 1247

\reference{}
Fan, X., et al. 2003, AJ, 125, 1649

\reference{}
Gnedin, N.\ Y. 2000, ApJ, 535, 530

\reference{}
Gnedin, N.\ Y. 2002, in Texas in Tuscany, eds.\ R.\ Bandiear, R.\ Maiolino,
F.\ Mannucci, (World Scientific: Singapore), 61

\reference{}
Gnedin, N.\ Y., \& Abel, T. 2001, NewA, 6, 437

\reference{}
Gunn, J.\ E., \& Peterson, B.\ A. 1965, ApJ, 142, 1633

\reference{}
Haiman, Z. 2002, 576, L1

\reference{}
Hu, E. M., Cowie, L.\ L., \& McMahon, R.\ G. 1998, ApJ, 502, L99

\reference{}
Hu, E. M., McMahon, R.\ G., \& Cowie, L.\ L. 1999, ApJ, 522, L9

\reference{}
Hu, E.\ M., Cowie, L.\ L., McMahon, R.\ G., Capak, P., Iwamuro, F., Kneib,
 J.-P., Maihara, T., \& Motohara, K. 2002, ApJ, 568, L75

\reference{}
Hu, E.\ M., Cowie, L.\ L., Capak, P., McMahon, R.\ G., Hayashino, T., \&
Komiyama, Y. 2004, AJ, 127, 563

\reference{}
Kodaira, K., et al. 2003, PASJ, 55, L17

\reference{}
Kneib, J.-P., Ellis, R.\ S., Santos, M.\ R., \& Richard, J. 2004, ApJ, in
press (astro-ph/0402319)

\reference{}
Liu, Y., Lin, J., Huang, G., Guo, Y., \& Duan, C. 2001, J.\ Opt.\ Soc.\
Am.\ B, 18, 666

\reference{}
Loeb, A., Barkana, R., \& Hernquist, L. 2004, ApJ, submitted
(astro-ph/0403193) 

\reference{}
Maier, C., et al. 2003, A\&A, 402, 79

\reference{}
Miralda-Escud\'e, J. 1998, ApJ, 501, 15
 
\reference{}
Pello, R., Schaerer,D., Richard, J., Le Borgne, J.-F., \& Kneib, J.,-P
2004, A\&A, 416, L35

\reference{}
Rhoads, J.\ E., et al. 2003, AJ, 125, 1006

\reference{}
Ricotti, M., Gnedin, N.\ Y., \& Shull, J.\ M. 2003, ApJ, 575, 49

\reference{}
Songaila, A. 2004, astro-ph/0402347

\reference{}
Spergel, D.\ N., et al. 2003, ApJ, in press (astro-ph/0302209)

\reference{}
Stanway, E.\ R., Glazebrook, K., Bunker, A.\ J., Abraham, R.\ G., Hook, I.,
Rhoads, J., McCarthy, P.\ J., Boyle, B., Colless, M., Crampton, D., Couch,
W., Jorgensen, I., Malhotra, S., Murowinski, R., Roth, K., Savaglio, S.,
\& Tsvetanov, Z. 2004, ApJ, 604, L13

\reference{}
Stern, D., \& Spinrad, H. 1999, PASP, 111, 1475

\reference{}
Taniguchi, Y., et al. 2003, ApJ, 585, L97

\reference{}
Weymann, R.\ J., Stern, D., Bunker,
A., Spinrad, H., Chaffee, F.\ H., Thompson, R.\ I., \& Storrie-Lombardi,
L. J. 1998, ApJ, 505, L95

\reference{}
White, R.\ L., Becker, R.\ H., Fan, X., \& Strauss, M.\ A. 2003, AJ,
126, 1

\end{references}
\end{document}